\documentclass[11pt,a4]{article}
\usepackage{amsthm,amsmath,latexsym,amssymb,amsfonts}
\usepackage{graphicx,lscape,fancyhdr,array,stmaryrd}

\newcommand{\bep}{\begin{picture}}
\newcommand{\eep}{\end{picture}}

\newcounter{YoungHeight}\newcounter{YoungWidth}

\newcounter{Mul1}\newcounter{Mul2}\newcounter{Mul3}\newcounter{Mul4}
\newcounter{A0}\newcounter{A1}\newcounter{A2}\newcounter{A3}\newcounter{A4}\newcounter{A5}\newcounter{A6}
\newcounter{B0}\newcounter{B1}\newcounter{B2}\newcounter{B3}
\newcounter{C1}\newcounter{C2}\newcounter{C3}\newcounter{C4}\newcounter{C6}\newcounter{C7}
\newcounter{D1}\newcounter{D2}\newcounter{D3}\newcounter{D4}\newcounter{D5}
\newcounter{T0}\newcounter{T1}

\newcounter{TGR0}
\newcounter{R0}\newcounter{R1}\newcounter{R2}\newcounter{R3}
\newcounter{AR0}\newcounter{AR1}\newcounter{AR2}\newcounter{AR3}\newcounter{AR4}\newcounter{AR5}\newcounter{AR6}\newcounter{AR7}
\newcounter{Dotted0}\newcounter{Dotted1}\newcounter{Dotted2}\newcounter{Dotted3}

\newcounter{reptA}
\newlength{\txtHShift}

\newlength{\txtWidth}

\newcommand{\HalfLength}[2]{\setcounter{Mul1}{#1}\setcounter{Mul2}{#1}\addtocounter{Mul1}{\value{Mul2}}\addtocounter{Mul1}{\value{Mul2}}%
\addtocounter{Mul1}{\value{Mul2}}\addtocounter{Mul1}{\value{Mul2}}\setcounter{#2}{\value{Mul1}}}

\newcommand{\Add}[3]{\setcounter{#1}{#2}\addtocounter{#1}{#3}}

\newcommand{\Length}[1]{#10}
\newcommand{\YoungScale}{}%\unitlength=0.35mm}

\newcommand{\shiftedText}[2]{{\hspace{#1}#2}}
\newcommand{\calcHShift}[1]{\settowidth{\txtWidth}{#1}\setlength{\txtHShift}{-0.5\txtWidth}}

\newcommand{\TextTop}[3]{{\calcHShift{#1}\HalfLength{#2}{T0}\Add{T1}{\Length{#3}}{-9}\put(\value{T0},\value{T1}){\shiftedText{\txtHShift}{#1}}}}

\newcommand{\BlockA}[2]{{\YoungScale\bep(\Length{#1},\Length{#2}){\Add{A1}{#1}{1}\Add{A2}{#2}{1}}%
\multiput(0,0)(10,0){\value{A1}}{\line(0,1){\Length{#2}}}\multiput(0,0)(0,10){\value{A2}}{\line(1,0){\Length{#1}}}%
\setcounter{YoungHeight}{\Length{#2}}\setcounter{YoungWidth}{\Length{#1}}\eep}}

\newcommand{\RectT}[3]{\bep(\Length{#1},\Length{#2})\put(0,0){\line(1,0){\Length{#1}}}\put(0,0){\line(0,1){\Length{#2}}}%
\put(\Length{#1},\Length{#2}){\line(-1,0){\Length{#1}}}\put(\Length{#1},\Length{#2}){\line(0,-1){\Length{#2}}}#3{#1}{#2}\eep}

\newcommand{\RectBRow}[4]{{\bep(\Length{#1},20)\put(0,0){\RectT{#2}{1}{\TextTop{#4}}}%
\put(0,10){\RectT{#1}{1}{\TextTop{#3}}}\eep}}

\newcommand{\YoungA}{\BlockA{1}{1}}
\newcommand{\YoungB}{\BlockA{2}{1}}

\newcommand{\YoungAA}{\BlockA{1}{2}}

\usepackage{amsthm,amsmath,latexsym,amssymb,amsfonts,amsbsy}
\usepackage{graphicx,lscape,fancyhdr,color,array,stmaryrd,euscript,wrapfig}
\usepackage{cite}

\newcommand{\be}{\begin{equation}}
\newcommand{\ee}{\end{equation}}
\newcommand{\bes}{\begin{split}}
\newcommand{\es}{\end{split}}
\newcommand{\bee}{\begin{eqnarray}}
\newcommand{\eee}{\end{eqnarray}}
\newcommand{\beee}{\begin{array}}
\newcommand{\bem}{\begin{multline}}
\newcommand{\eem}{\end{multline}}
\newcommand{\bec}{\begin{Comment}}
\newcommand{\ec}{\end{Comment}}

\newcommand{\Y}[1]{{\ensuremath{\mathbb{Y}(#1)}}}

\newcommand{\Yy}{\ensuremath{\mathbf{Y}}}
\newcommand{\Xx}{\ensuremath{\mathbf{X}}}

\newcommand{\DO}{{D_{\Omega}}}
\newcommand{\smallpic}[1]{{\unitlength=0.2mm#1}}

\newcommand{\boldpic}[1]{{\linethickness{0.4mm}#1}}

\newcommand{\fud}[2]{{^{#1}_{\phantom{#1}#2}}}
\newcommand{\fdu}[2]{{_{#1}^{\phantom{#1}#2}}}
\newcommand{\fdud}[3]{{_{#1\phantom{#2}#3}^{\phantom{#1}#2}}}

\newcommand{\fudud}[4]{{^{#1\phantom{#2}#3}_{\phantom{#1}#2\phantom{#3}#4}}}

\definecolor{rougef}{rgb}{0.56,0,0}
\definecolor{vertf}{rgb}{0,0.5,0}
\definecolor{bleuf}{rgb}{0,0,0.8}

\begin{document}
\begin{titlepage}

\begin{flushright}
\vspace{1mm}
%FIAN
\end{flushright}

\vspace{1cm}
\begin{center}
{\bf \Large  Non-abelian cubic vertices for higher-spin fields in $AdS_d$}
\vspace{2cm}

\textsc{Nicolas Boulanger\footnote{Research Associate of the Fund
for Scientific Research-FNRS (Belgium);
nicolas.boulanger@umons.ac.be}, Dmitry
Ponomarev\footnote{dmitri.ponomarev@umons.ac.be} and
E.D. Skvortsov\footnote{skvortsov@lpi.ru}}

\vspace{2cm}

{\em${}^{1,2}$  Universit\'e de Mons -- UMONS, 20 Place du Parc, 700 Mons, Belgium}

{\em${}^3$  Albert Einstein Institute, Golm, Germany, D-14476, Am M\"{u}hlenberg 1}

{\em${}^3$  Lebedev Institute of Physics, Moscow, Russia, 119991, Leninsky pr-t, 53}

\end{center}

\vspace{0.5cm}
\begin{abstract}

We use the Fradkin-Vasiliev procedure to construct the full set of non-abelian
cubic vertices for totally symmetric higher spin gauge fields in $AdS_d$ space.
The number of such vertices is given by a certain tensor-product multiplicity.
We discuss the one-to-one relation between our result and the list of non-abelian gauge
deformations in flat space obtained elsewhere via the cohomological approach.
We comment about the uniqueness of Vasiliev's simplest higher-spin algebra in relation
with the (non)associativity properties of the gauge algebras that we classified.
The gravitational interactions for (partially)-massless (mixed)-symmetry fields are also discussed.
We also argue that those mixed-symmetry and/or partially-massless fields that are described by
one-form connections within the frame-like approach can have nonabelian interactions among themselves and again the number of nonabelian vertices should be given by tensor product multiplicities.
\end{abstract}

\end{titlepage}

\numberwithin{equation}{section}

\section{Introduction}
\label{intro}

In 1987, Fradkin and Vasiliev solved the problem of cubic interactions for
higher spin gauge fields among themselves and with gravity \cite{Fradkin:1986qy,Fradkin:1987ks}.
A key ingredient of their construction was the expansion of all fields around
the anti-de Sitter ($AdS$) background of dimension four instead of the Minkowskian
flat background.
It proved very convenient to use commuting $sl(2,\mathbb{C})$ Weyl spinors
although there is \emph{a priori} nothing fundamental, at that stage, with
the dimensionality of the background. The problem can indeed be considered
in an $AdS_d$ background of arbitrary dimension $d>4\,$, which is the framework
of the present paper where we explicitly build and classify all the possible
non-abelian couplings between totally-symmetric higher-spin (including spin-$2$)
gauge fields in $AdS_d\,$ with $d>4\,$.
By non-abelian cubic vertices, we mean those which non-trivially
deform the abelian gauge algebra of the free theory.
%\vspace*{.1cm}

For that purpose, we use the Fradkin-Vasiliev procedure whereby the
free theory is presented in the frame-like approach, starting from a Lagrangian
quadratic in the linear curvature two-forms.
The cubic vertices are obtained by substituting non-linear deformations
of the curvature two-forms inside the quadratic, free action.
The very structure of these non-linear deformations automatically
implies that the gauge algebra is non-abelian to the first non-trivial
order in deformation.
We also adopt the MacDowell-Mansouri-Stelle-West formulation
of gravity \cite{MacDowell:1977jt,Mansouri:1977ej,Stelle:1979aj}
and its higher-spin generalization \cite{Vasiliev:2001wa}
which makes the $AdS_d$ symmetry manifest through the introduction of
an extra field, sometimes called compensator, inside the Lagrangian.
For recent works along the same lines, see e.g.
\cite{Alkalaev:2010af,Zinoviev:2010av,Boulanger:2011qt,Zinoviev:2011fv,Boulanger:2011se,Vasiliev:2011xf}.
\vspace*{.1cm}

Our main result can be stated in a concise way: given three totally-symmetric gauge
fields with spins $s$, $s'$ and $s''$, the number of independent non-abelian vertices
is given by the tensor product multiplicity
\be \label{tworowmul}\parbox{42pt}{\boldpic{\RectBRow{4}{4}{$s-1$}{}}}\bigotimes
\parbox{50pt}{\boldpic{\RectBRow{5}{5}{$s'-1$}{}}}\longrightarrow
\parbox{60pt}{\boldpic{\RectBRow{6}{6}{$s''-1$}{}}}\,,  \ee
\emph{i.e.} by all the possible independent ways to contract two rectangular two-row
$so(d-1,2)$ tensors in order to form another two-row rectangular $so(d-1,2)$ tensor
of given length.
The lengths of the diagrams involved are related to the spins as indicated above. The gauge fields valued
in such irreducible tensor representations of the anti-de Sitter algebra $so(d-1,2)$ have been proposed for
the description of higher-spin fields by Vasiliev in \cite{Vasiliev:2001wa}.
At the same time, this multiplicity equals the number of non-abelian vertices in Minkowski
space, \cite{Bekaert:2010hp}.
The vertices we construct are off-shell and not subjected to any transverse/traceless gauge condition. A particular way of contracting indices in (\ref{tworowmul}) is given by the Vasiliev higher-spin algebra \cite{Vasiliev:2003ev}. This algebra is a unique associative algebra with spectrum of generators (\ref{tworowmul}) and all other non-abelian deformations lead to nonassociative algebras that can hardly be consistent at the quartic level.

Among all the possible types of vertices: abelian, non-abelian, Chern-Simons-like
etc.\footnote{See e.g. \cite{Vasiliev:2011xf} for some terminology in the present context,
and \cite{Barnich:2000zw} in the general case of a local gauge theory.},
the non-abelian ones contain more information about the full theory, whatever it is.
Consistency at the quartic level may however require some abelian cubic vertices to be added,
see the discussion in \cite{Boulanger:2008tg} and \cite{Bekaert:2010hw}.

The construction that we present here
for the classification of the non-abelian cubic vertices in $AdS_d$ uses the
$sp(2)$ technology developed by Vasiliev and collaborators
\cite{Vasiliev:2003ev,Vasiliev:2004cm,Alkalaev:2007bq}
and shows some similarity with the cohomological method \cite{Barnich:1993vg} used
in \cite{Bekaert:2010hp} for the classification of the non-abelian algebras
in flat space, to first order in deformation.
Both the present approach to consistent vertices and the cohomological one
have the advantage that they provide a completely algebraic
reformulation of the consistent-coupling problem.
It is \emph{a priori} not clear that the Fradkin-Vasiliev ansatz
leads to the most general non-abelian deformations. As a matter of fact,
and in agreement with what was argued in \cite{Vasiliev:2011xf},
we find that it actually produces the exhaustive list of non-abelian
cubic vertices in $AdS_d\,$.
This follows from the following argument:
On the one hand, we have at our disposal \cite{Bekaert:2010hp}
the complete classification of non-abelian gauge-algebra deformations,
for any given triplet $(s,\,s',\, s^{''})$
of higher-spin gauge fields in flat background.
On the other hand we know that
to every non-abelian vertex in $AdS_d$ for totally symmetric gauge fields
there is  a corresponding non-abelian vertex in flat space \cite{Boulanger:2008tg}.
Therefore, if one constructs -- as we do in this paper --
a list of independent non-abelian vertices in $AdS_d\,$
whose number corresponds to the number of non-abelian vertices in flat space,
then one automatically has access to the full list of non-abelian vertices in $AdS_d\,$.
Indeed, assuming the existence of additional, independent non-abelian vertices in $AdS_d\,$,
the corresponding flat limit along the lines of \cite{Boulanger:2008tg} --
which entails starting from the nontrivial terms in the Lagrangian
containing the highest number of partial derivatives, a filtration that can always
be done for cubic vertices in $AdS_d$ --
would give rise to additional and independent non-abelian vertices in flat space,
thereby giving a total number of non-abelian vertices exceeding the upper bound
%of possible non-abelian deformations classified
obtained in \cite{Bekaert:2010hp}.
\vspace*{.1cm}

Manifestly covariant cubic vertices in flat space of arbitrary dimension
have been explicitly written by many authors by now
\cite{Sagnotti:2010at,Fotopoulos:2010ay,Manvelyan:2010je,Metsaev:2012uy}\footnote{The fundamental
results on cubic interaction have been obtained by Metsaev within the light-cone approach,
\cite{Metsaev:1993ap, Metsaev:1993mj,Metsaev:2005ar}. For
a non-technical review on higher-spin gravity that includes
a discussion on cubic vertices, see \cite{Bekaert:2010hw}. See also \cite{Sagnotti:2011qp}.}.
The situation is not exactly the same in $AdS_d\,$, see e.g. \cite{Bekaert:2010hk,Joung:2012fv,Manvelyan:2012ww}
for some very recent endeavours. A noticeable exception is the very general analysis
provided in \cite{Vasiliev:2011xf}, that shows how to classify vertices
in $AdS_d$ using the frame-like formalism. In \cite{Vasiliev:2011xf} the set of generating
functions for non-abelian vertices has also been suggested.
Our goal is to elaborate on the algebraic structure of non-abelian cubic vertices.
Vertices that explicitly involve the (generalized) Weyl tensors
will not be studied here.
The triplets of spins $(s, s', s'')$ with
$s\leqslant s' \leqslant s^{''}$ considered in \cite{Vasiliev:2011xf} have to satisfy
the triangle inequality $s^{''} < s + s'\,$ that coincides with the necessary condition
obtained in \cite{Bekaert:2010hp} for the existence of non-abelian vertices in flat spacetime.
\vspace*{.2cm}

We also discuss gravitational interactions of various (partially)-massless (mixed)-symmetry
fields. The gravitational interactions are the simplest ones and we show
that these can always be introduced for certain types of gauge fields, though not for all
interestingly enough. At the end we give a general argument that the number of nonabelian vertices among
various (partially)-massless (mixed)-symmetry fields should again be given by certain tensor product multiplicities.
\vspace*{.2cm}

The plan of the paper is as follows. Section \ref{Free}
reviews the frame-like formulation of free, totally symmetric
higher-spin gauge fields in manifestly $so(d-1,2)$-covariant fashion
along the lines of \cite{Vasiliev:2001wa}.
In Section \ref{FV} we briefly review the
Fradkin-Vasiliev ansatz for cubic, non-abelian
vertices in $AdS_d\,$, in the frame-like formalism.
A more detailed account can be found in \cite{Vasiliev:2011xf}.
In Section \ref{star} we present the $sp(2)$-invariant operators
from which we construct the full list of non-abelian gauge algebras
for candidate cubic vertices.
In Section \ref{star} we also show that, among
the various gauge algebras that are obtained
at the first nontrivial order in interaction,
only one can be elevated to an associative,
infinite-dimensional higher-spin algebra.
This algebra is nothing but the algebra
originally found by Eastwood \cite{Eastwood:2002su},
isomorphic to the one used by Vasiliev \cite{Vasiliev:2003ev}
for his construction of fully nonlinear equations in $AdS_d\,$.
We then show in Section \ref{tac}  that all the possible, non-equivalent
gauge algebra deformations are indeed realized
by consistent cubic vertices, and that their number coincides
with the total number of non-abelian gauge algebras in
flat spacetime found in \cite{Bekaert:2010hp}.
The computation of some coefficients
entering the cubic vertices is given in
the Appendix.
The gravitational interactions for more general types of fields
including partially-massless fields and mixed-symmetry fields
are considered in Section \ref{sec:gravitational}.
Finally, Section \ref{sec:conclusions} contains our conclusions.

%===================================================
\section{Free fields and the linear action}
\label{Free}
%===================================================

Nonlinear equations for an infinite tower of
totally symmetric gauge fields in arbitrary
dimension have been given by Vasiliev in \cite{Vasiliev:2003ev}.
These equations are background independent,
but the gauge algebra contains the $AdS_d$ algebra
as maximal finite-dimensional subalgebra,
and the simplest exact solution of Vasiliev's
equations is empty $AdS_d$ spacetime.

The $AdS_d$ exact solution around which one can
linearize the full nonlinear
equation is presented in the way used by
MacDowell-Mansouri and Stelle-West \cite{MacDowell:1977jt,Stelle:1979aj}:
\begin{equation}
\label{ads}
R_0^{A,B}=(D_0)^2={\rm d}W_0^{A,B}+W_0^{A,C}\wedge W_{0C}{}^B=0,
\end{equation}
where $W_0^{A,B}=-W_0^{B,A}$
is the background 1-form connection transforming in the adjoint
representation of $so(d-1,2)\,$,
namely in the antisymmetric rank-2 representation
of  $so(d-1,2)\,$.
The differential $D_0$ is the corresponding
covariant derivative around $AdS_d\,$.
The important
ingredient that allows to combine the vielbein and spin-connection fields of Lorentz-covariant
formulation of gravity into the single
$so(d-1,2)\,$-connection
$W_0$ is the compensator vector $V^A$ that is
constrained to satisfy
\begin{equation}
\label{comp}
V^AV^B\eta_{AB} =-\Lambda^{-1},
\end{equation}
where $\Lambda$ is the cosmological constant.
Assuming one fixes $V$ in such a way that
(\ref{comp}) is satisfied,
the algebra of $so(d-1,2)$ rotations
preserving $V$ is identified with the Lorentz
algebra $so(d-1,1)\,$.
Then one can introduce a one-form frame field
\begin{equation}
\label{frame}
E_0^A := D_0V^A={\rm d}V^A+W_0{}^A{}_BV^B,
\end{equation}
which is assumed to have maximal rank $d\,$.
From (\ref{comp}) we find
\begin{equation}
\label{ev}
E_0^AV_A=0\;.
\end{equation}

A spin-$s$ massless field in $AdS_d$ spacetime
can be  described \cite{Vasiliev:2001wa}
by a one-form
$W^{A(s-1),B(s-1)}$
carrying the
irreducible representation of the $AdS_d$ isometry algebra $so(d-1,2)$
described by the traceless two-row rectangular Young diagram of length $s-1$.
Then one constructs the linearized higher-spin two-form curvature
\begin{equation}
\label{hslincur}
R_1^{A(s-1),B(s-1)}=D_0W^{A(s-1),B(s-1)}\;.
\end{equation}
The curvature (\ref{hslincur}) is gauge
invariant with respect to abelian gauge transformations
\begin{equation}
\label{hsgaugetr}
\delta_0 W^{A(s-1),B(s-1)}=D_0\xi^{A(s-1),B(s-1)},
\end{equation}
which follows from the fact that
$(D_0)^2 = 0\,$.

To properly describe free massless spin-$s$ field
one should impose the following equations of
motion \cite{Lopatin:1987hz,Vasiliev:2001wa}, called the first on-mass-shell theorem,
\begin{equation}
\label{fot}
R_1^{A(s-1),B(s-1)}\approx E_{0}^ME_{0}^N
C^{A(s-1)}{}_{M,}{}^{B(s-1)}{}_N\;,
\end{equation}
where $C^{A(s),B(s)}$ is an irreducible
two-row $so(d-1,2)$ tensor
subjected to the extra $V$-transversal constraint
\begin{equation}
\label{vtrans}
C^{A(s-1)M,B(s)}V_M=0\;.
\end{equation}
The zero-form $C^{A(s),B(s)}$
generalizes the Weyl tensor
of gravity to the higher-spin case, in the sense
that, in the spin-2 case, the Einstein equations
linearized around AdS can be written in the form
\begin{equation}
\label{Einstein}
R_1^{A,B} \approx E_{0}^ME_{0}^N
C^{A}{}_{M,}{}^{B}{}_N\;,
\end{equation}
where $C^{A(2),B(2)}$ only contains the
linearized Weyl tensor of gravity when decomposed
under~${so}(d-1,1)\,$, as a consequence
of the $V$-transversality condition
(\ref{vtrans}).

The quadratic action for the symmetric spin-$s$ gauge field is \cite{Vasiliev:2001wa}
\begin{equation}
\label{linaction}
S_0=\frac{1}{2}\int d^dx\sum_{p=0}^{s-2}a(s,p)
V_{C(2(s-2-p))}G_{MNPQ}{R_1}^{MB(s-2),NC(s-2-p)D(p)}
{R_1}^P{}_{B(s-2),}{}^{QC(s-2-p)}{}_{D(p)},
\end{equation}
where
\begin{equation*}
V_{A(n)}=\overbrace{V_A\dots V_A}^n, \qquad
G_{M_1M_2\dots M_i}=\epsilon_{NM_1M_2\dots M_iR_{i+1}\dots R_{d}}
V^NE_0^{R_{i+1}}\dots
E_0^{R_{d}},
\label{notation1}
\end{equation*}
\begin{equation*}
a(s,p)=\alpha_s(-1)^p\Lambda^{-p}\frac{(d-5+2(s-p-2))!!(s-p-1)}{(d-5)!!(s-p-2)!}
\end{equation*}
and $\alpha_s$ is an arbitrary normalization coefficient.

%==============================================
\section{Fradkin-Vasiliev procedure}
\label{FV}
%===============================================

\paragraph*{Deformation procedure.}

Given a quadratic action $S_0$ (\ref{linaction})
that is gauge invariant under
the gauge transformation
 (\ref{hsgaugetr}) one looks for a deformation
of both the action and gauge transformations by higher-order, field-dependent corrections $S=S_0+g\,S_1+{\cal O}(g^2)\,$, $\delta=\delta_0+g\,\delta_1+{\cal O}(g^2)\,$.
The consistency condition reads
\begin{align}
\delta_0S_0+ g\left(\delta_1 S_0+\delta_0 S_1\right)+g^2\left(\delta_1 S_1+\delta_0 S_2+\delta_2 S_0\right)+{\cal O}(g^3)=0
\end{align}
with the first term vanishing because of gauge invariance of $S_0\,$.
At the cubic level one looks for a solution
of $\delta_1 S_0+\delta_0 S_1 =0 \,$.
If one succeeds in finding such a cubic part
$S_1$ whose variation under linearized gauge transformations $\delta_0$ vanishes on free mass-shell,
then it implies that $\delta_0 S_1$ is proportional to free field equations
\begin{eqnarray}\label{cubicinv}
\delta_0 S_1 &=&
F\left(\frac{\delta S_0}{\delta W}, \xi, W\right)\;,\quad {\rm and} \quad
F(0,\xi, W)~=~0\,,
\end{eqnarray}
where $F$ is trilinear in its arguments and can
be used to extract $\delta_1\,$.
As always, the cubic action $S_1$ and the gauge
transformations $\delta_1$ are defined modulo
field and gauge parameter redefinitions.
The problem of extracting $\delta_1$ out of $F$ is
purely technical and one does not need to solve it
once a nontrivial solution to $S_1$ is found.

The Fradkin-Vasiliev procedure
\cite{Fradkin:1987ks,Fradkin:1986qy} does not
give the general solution to the problem of
constructing of cubic action. However, as we will
show below, it actually leads to the exhaustive
list of non-abelian cubic vertices. To cover all cubic vertices
one has to extend the Fradkin-Vasiliev setup with Weyl zero-forms,
see \cite{Vasiliev:2011xf} for more detail.

\paragraph*{Yang--Mills-like transformations.}

The Fradkin-Vasiliev \cite{Fradkin:1987ks,Fradkin:1986qy}
procedure is based on the idea that one should
look for Lagrangians that are quadratic in
the curvature two-forms, similarly to what
happens in Yang-Mills theory.
In other words, in order to generate
a cubic action, one replaces the linearized
curvature $R_1$ by a nonlinear completion $R_2$
of it inside the quadratic action
$S\sim \int R_1 R_1$
from which one starts.
Indeed, the action (\ref{linaction}) is quadratic
in the curvatures.

This implies that the one-forms
$\{W^k\}$ are valued in some internal algebra
whose product we denote by $\diamond\,$,
with the understanding that the algebra
is not necessarily associative.
To fix the notation, we have
$T_m\,\diamond \,T_n = \tfrac{1}{2}\,g^k_{mn}T_k\,$,
where the $T$'s give
a basis of the (possibly non-associative)
internal algebra ${\cal A}$
to which the one-forms belong.
As we are going to construct the most
general non-abelian cubic vertices coupling
symmetric gauge fields around $AdS_d$,
the symbol $\diamond$ does not denote the
star product of Vasiliev's theory.
As we said,
it denotes an arbitrary product that acts on
rectangular Young diagrams of
$so(d-1,2)$ and can be non-associative.
 The curvature
\begin{equation}
\label{YMcurv}
R={\rm d}W + W \diamond W
\end{equation}
is given, in components along the generators
$T_k$, as
\begin{equation}
\label{YMcurvcomp}
R^k= {\rm d}W^k  + f^k_{mn}\, W^m\wedge W^n  \;,
\quad f^k_{mn}:=g^k_{[mn]}\;.
\end{equation}
Under the Yang--Mills-like gauge transformation
\begin{equation}
\label{YMgauge}
  \delta^{YM} W= {\rm d} \xi + [W, \xi]_\diamond \;,
\end{equation}
the curvature transforms as
\begin{equation}
\label{curvtr}
 \delta^{YM} R = [R,\xi]_\diamond + JAC
\end{equation}
where
\begin{eqnarray}
JAC &:=&\left[ \xi\diamond(W\diamond W) - (\xi\diamond
W)\diamond W\right]
- \left[ W\diamond(\xi\diamond W) - (W \diamond \xi)\diamond W\right]
\\
&& +  \left[ W\diamond(W\diamond\xi) - (W \diamond W)\diamond \xi\right]
\end{eqnarray}
is the Jacobiator. It vanishes for an associative
algebra. We will not be bothered by the Jacobiator
in the following, since it comes at order
$W^ 2$ and for the problem of cubic vertices
we only need the transformation
of the curvature to order $W\,$.
As it will be shown below,
to achieve gauge invariance of the cubic vertices,
the gauge transformation will receive an extra
piece $\delta_1^{ext}$
having no simple geometrical interpretation
in the current framework.
However, as long as we are interested in the
cubic vertices and not in the explicit form of
 $\delta_1^{ext}W\,$, this issue will not be
 relevant to us.

\paragraph*{Perturbation around $AdS_d$.}

We want to include the $AdS_d$ connection
as part of the set of one-forms $W^k\,$,
or in other words,
we include the $so(d-1,2)$ generators
among the generators $T_k$
of the internal algebra ${\cal A}\,$.
We ask that the one-form gauge fields should be
expanded around the $AdS_d$ background solution (\ref{ads})
\begin{equation}
\label{ads1}
R_0={\rm d}W_0+W_0 \diamond W_0=0\;,
\end{equation}
namely, we have the weak field decomposition
$W=W_0 + W_1\,$ and impose a constraint
on the $\diamond\,$-product
among the class of two-row Young
tableaux with a single column, \parbox{12pt}{\YoungAA},
namely, that (\ref{ads1}) should be identical with (\ref{ads}).
Then, the curvature (\ref{YMcurv}) reads,
to first order in expansion around $AdS_d\,$,
as
\begin{equation}
\label{hslincur1}
R_1 = {\rm d} W + W_0 \diamond W+W \diamond W_0\;.
\end{equation}
Again, we impose that this formula should reproduce (\ref{hslincur}),
which gives an additional restriction on
the $\diamond\,$-product and implies that the
higher spin fields $W_1$ transform
as tensors under the (adjoint) action of $so(d-1,2)\subset {\cal A}\,$.
The linearized gauge transformation (\ref{hsgaugetr}) reads
\begin{equation}
\label{hsgaugetr1}
\delta_0 W = {\rm d}\xi + W_0\diamond \xi - \xi \diamond W_0\;,
\end{equation}
and we rewrite the quadratic action (\ref{linaction})
in the form
\begin{equation}
\label{linaction1}
S^{\{s\}}_0[W^s]=\int \,\langle R^{\{s\}}_1\,,\,
R^{\{s\}}_1\rangle_{W_0} \;,
\end{equation}
where we added a label $\{s\}$ in order to specify the spin under consideration.

\paragraph{Cubic ansatz.}
At the next stage we seek a cubic
deformation of the quadratic Lagrangian.
Following Fradkin and Vasiliev, the idea is to
keep the form of the quadratic action
(\ref{linaction1})
and replace the linear curvature $R_1$ with the
non-linear $R=R_1+R_2\,$, where
\begin{equation}
\label{curvcub}
R_2=W\diamond\, W\;,
\end{equation}
so as to obtain
\begin{equation}
\label{cubaction}
S_0+S_1 + {\cal O}(W^4) =\sum_s \alpha_s \int
\langle R^{\{s\}}\,,\,
 R^{\{s\}}\rangle_{W_0} \;.
\end{equation}
We want to constrain the $\diamond$-product  in
such a way that
$\delta_1^{YM} S_0+\delta_0 S_1$ should vanish
on the free shell, up to terms of order
${\cal O}(W^3 \xi)\,$, where
$\delta^{YM}_1$ is the part of (\ref{YMgauge})
that is linear in the weak fields:
\begin{equation}
\label{nagtr}
\delta_1^{YM} W= W\diamond \xi
-\xi \diamond W\;.
\end{equation}
Taking into account that $\delta_0 R_2+\delta_1^{YM} R_1=[R_1,\xi]_{\diamond}\,$
(non-associative terms in (\ref{curvtr}) do not contribute at this order),
 one can easily compute the variation of the action:
\begin{eqnarray}
\delta_1^{YM} S_0+\delta_0 S_1 &=& 2
\sum_s\alpha_s\int \langle
R^{\{s\}}_1 \, , \, [R_1,\xi]^{\{s\}}_{\diamond}
\rangle_{W_0}
+{\cal O}(W^3\xi),
\end{eqnarray}
where $ [R_1,\xi]^{\{s\}}_{\diamond}$ denotes
the restriction
of $ [R_1,\xi]_{\diamond}$ to the spin-$s$ sector.
According to the central result recalled in
(\ref{fot}),
this variation on free shell gives
\begin{align}
\label{FVcond}
\delta^{YM}_1 S_0+\delta_0 S_1 \;\approx\;
2\sum_s\alpha_s\int \langle (E_0E_0 C)^{\{s\}}
\,,\, ([E_0E_0 C,\xi]_{\diamond})^{\{s\}}\rangle_{W_0}
+{\cal O} (W^3\xi )\;,
\end{align}
where $(E_0E_0 C)$ is the {\it r.h.s.} of (\ref{fot}).

By arguments similar to those at the beginning of
this section, if one succeeds in adjusting the
free coefficients $\alpha_s$
in such a way that the gauge variation
(\ref{FVcond}) is zero on free shell and
up to terms cubic in the fields,
then there exists a certain completion $\delta_1^{ext}$
of $\delta_1^{YM}$ that yields the full gauge invariance of the
action $S_0+S_1\,$.
We recall that $\delta_1$ was split into a
Yang--Mills-like part $\delta^{YS}_1$
plus the rest $\delta_1^{ext}\,$,
where the latter
 cannot be presented in a simple, geometric, form
 within the current approach.
Whatever $\delta_1^{ext}\,$ is, the vanishing
of the right-hand side of (\ref{FVcond})
up to terms cubic in the fields is sufficient
to prove that the action $S_0+S_1\,$
is gauge invariant under a certain
$\delta_1$ transformation containing the
non-abelian part $\delta_1^{YM}\,$.
Let us add the comment that, by construction,
$\delta^{ext}_1W$ is linear in $R_1$ and in
the gauge parameters, and therefore does not
contribute to the non-abelian nature of the
gauge transformation at the first nontrivial
order where we work; only $\delta_1^{YM}$ does.
\vspace*{.3cm}

The Fradkin-Vasiliev procedure amounts to solving
 \begin{align}
\label{FVcond1}
0~=~\sum_{k,m,n}\alpha_k\int \langle (E_0E_0 C)^{\{k\}}
\,,\,
f^k_{mn}
(E_0E_0 C)^{\{m\}} \xi^{\{n\}} \rangle_{W_0}
=: \sum_{k,m,n} I^k_{mn}
\end{align}
for the free coefficients $\alpha_s\,$ and
for the structure constants $f^k_{mn}\,$.
Let us consider the terms in (\ref{FVcond1})
involving only fields and
gauge parameters of three fixed spins $k$, $m$ and $n\,$.
Obviously, such terms are independent from the others
and to solve (\ref{FVcond1})
they should cancel among each other.
The Fradkin--Vasiliev condition, in the fixed
sector we consider, therefore reads
\begin{equation}
\label{FVcond2}
I^k_{m,n}+I^m_{n,k}+I^n_{k,m} = 0 \;.
\end{equation}
Regrouping terms pairwise,
it implies that one should have
\begin{equation}
 \label{FVcond3}
 \alpha_kf^k_{nm} \int \langle (E_0E_0C)^{\{k\}}
\,,\, \xi^{\{n\}} (E_0E_0C)^{\{m\}}\rangle_{W_0}
= \alpha_m f^m_{kn} \int \langle
(E_0E_0C)^{\{m\}} \,,\,
(E_0E_0C)^{\{k\}} \xi^{\{n\}}\rangle_{W_0} \;,
 \end{equation}
where there is no sum over the Latin indices
$k,m$ and $n\,$.
Our aim in this paper is therefore
to find the most general solution of the above equation.
\vspace*{.3cm}

\noindent This leads us to the following two problems:
 \begin{itemize}
 \item[1.] Find the full set of independent
 $f^k_{mn}$ coefficients.
 This is done in the next
 Section~\ref{star};
 \item[2.] For each independent product rule
 found in item 1, solve  (\ref{FVcond3}).
 This is done in Section \ref{tac}.
 \end{itemize}

\section{Non-abelian deformations}
\label{star}

Let us recall that the one-form gauge fields
$W^s$ entering the formulation of free
higher-spin theory around $AdS_d$ transform
as $so(d-1,2)$ tensors characterized
by a Young diagram made of two rows of equal
lengths $(s-1)\,$. In the spin-s sector one therefore has
the following correspondence
\begin{eqnarray}
W^s \leftrightsquigarrow W^{A(s-1),B(s-1)}\;.
\end{eqnarray}
It is convenient to follow the notation of
\cite{Vasiliev:2001wa,Vasiliev:2003ev}
and introduce a set of $2(d+1)$ bosonic
oscillators $Y^A_\alpha\,$, $\alpha=1,2\,$
that are used to realize $\mathfrak{sp}(2)$ generators
$K_{\alpha\beta}$ by
\begin{equation}
K_{\alpha\beta} := \frac{{\rm i}}{2}\;\left( Y^A_\alpha
\frac{\partial}{\partial Y^{\beta\, A}} +
Y^A_\beta
\frac{\partial}{\partial Y^{\alpha\, A}}\right) \;,
\end{equation}
so that indeed
\begin{equation}
[K_{\alpha\beta}\,,\,K_{\gamma\delta}] =
\epsilon_{\gamma(\alpha}K_{\beta)\delta}+
\epsilon_{\delta(\alpha}K_{\beta)\gamma}\;,
\end{equation}
where one raises and lowers indices with the
$\mathfrak{sp}(2)$-invariant symbol
$\epsilon^{\alpha\beta}=-\epsilon^{\beta\alpha}$
according to the rule
$Y^\alpha = \epsilon^{\alpha\beta}Y_\beta\,$,
$Y_\alpha = Y^\beta\epsilon_{\beta\alpha}\,$ where
$\epsilon^{12}=1=\epsilon_{12}\,$.
One then represents the spin-s gauge field by
\begin{equation}
W^s :=  \tfrac{1}{(s-1)!(s-1)!}\;
W^{A(s-1),B(s-1)} Y_A^1\ldots Y_A^1\;
Y_B^2\ldots Y_B^2
\end{equation}
so that the $\mathfrak{sp}(2)$-singlet conditions
\begin{equation}
[K_{\alpha\beta}\,,\,W^s] = 0\;
\end{equation}
impose that the coefficients $W^{A(s-1),B(s-1)}$
are two-row irreducible tensors of
$\mathfrak{gl}(d+1)\,$, see e.g.
\cite{Vasiliev:2011xf} for more details and
references.
\vspace*{.3cm}

Given two $\mathfrak{sp}(2)$-singlet fields $W^n(Y)$ and $W^m(Z)\,$
-- we hereby double
the set of $Y^A_\alpha$ oscillators by introducing
the oscillators $Z^A_\alpha$ that play exactly the same
role, there is
a natural operator that contracts a pair of indices:
\begin{align*}
\tau^{\alpha\beta}_{YZ}&:=\frac{\partial^2}{\partial Y^A_{\alpha}\partial Z^{\phantom{A}}_{A\beta}}\;.
\end{align*}
{}From $W^n(Y)$ and $W^m(Z)\,$ one can
produce another $\mathfrak{sp}(2)$ singlet by
acting on the product $W^n(Y) \,W^m(Z)\,$ with
some $\mathfrak{sp}(2)$-invariant operator
built out of $\tau^{\alpha\beta}_{YZ}\,$
and then setting $Z^A_\alpha=Y^A_\alpha\,$.
As an  $\mathfrak{sp}(2)$ module,
$\tau_{\alpha\beta}$ decomposes into
$\bullet\oplus\smallpic{\YoungB}\;$,
so that the problem is to find all the
$\mathfrak{sp}(2)$-invariants of
$\bullet\oplus\smallpic{\YoungB}\,$.
There are two generating
$\mathfrak{sp}(2)$-invariants:
\begin{eqnarray}
s_{YZ}&:=& \tau^{\alpha\beta}_{YZ}
\epsilon_{\alpha\beta}
\;\equiv\; \frac{\partial^2}{\partial Y^A_{1}\partial Z^{\phantom{A}}_{2A}}-
\frac{\partial^2}{\partial Y^A_{2}\partial Z^{\phantom{A}}_{1A}} \;,
\\
 p_{YZ} &:=& \det{(\tau^{\alpha\beta}_{YZ})}
\; \equiv\; \frac{\partial^2}{\partial Y^A_{1}\partial Z^{\phantom{A}}_{1A}}\frac{\partial^2}{\partial Y^B_{2}\partial Z^{\phantom{B}}_{2B}}-\frac{\partial^2}{\partial Y^A_{1}\partial Z^{\phantom{A}}_{2A}}\frac{\partial^2}{\partial Y^B_{2}\partial Z^{\phantom{B}}_{1B}}\;.
 \label{sandt}
\end{eqnarray}

The Vasiliev higher-spin algebra
\cite{Vasiliev:2003ev} is defined as a
certain quotient of the Weyl algebra or of the
universal enveloping
algebra ${\cal U}(so(d-1,2))\,$,
where the Weyl algebra is realized by the
star product algebra
\begin{eqnarray}
\label{starproduct}
W^n(Y)\star W^m(Y) &=& \left. \exp{\left(\tfrac{1}{2}\,s_{YZ}\right)}W^n(Y) W^m(Z)\right|_{Z=Y}
\end{eqnarray}
modulo the ideal generated by the traces.
Fortunately for us, the long tail of terms projecting out the ideal
does not contribute to the Fradkin-Vasiliev
condition (\ref{FVcond3}) since any
$\eta^{AB}$-proportional
term vanishes in the variation of the action
when put on the free mass-shell.

One may try to define some other
$\mathfrak{sp}(2)$-invariant product rules
via
\begin{align}
W^n(Y)\diamond W^m(Y) &= \left.\sum_k  \chi^k_{n,m}(s_{YZ}, p_{YZ}) W^n(Y) W^m(Z)\right|_{Z=Y},
\label{diamond}
\end{align}
with $\chi^k_{n,m}(\bullet, \bullet)$ being a
polynomial function in its two arguments, which
can depend on $k,m$ and $n\,$.
The list of all the possible inequivalent
functions
$\chi^k_{n,m}$ gives all the inequivalent
ways to contract two two-row $so(d-1,2)$-Young
diagrams with lengths $\check{n}:=n-1$ and
$\check{m}$ in order to produce a similar Young
diagram with length $\check{k}\,$.
The corresponding composition rules
(\ref{diamond}) are not associative
if they contain at least one $p$ operator.
Indeed, by the universal property, the only associative
algebra on the vector space of two-row rectangular
$so(d-1,2)$ Young tableau
is given by
${\cal A} \cong
\frac{{\cal U}(so(d-1,2))}{I_{singl.}}\;$,
where $I_{singl.}$ is the ideal that annihilates
the scalar Dirac singleton, see e.g.
\cite{Iazeolla:2008ix} and references therein.
The corresponding associative product (\ref{starproduct})
is generated by the $s$ contraction only.
Crucial in this line of reasoning is the fact
that the higher-spin tensors generating the
algebra under consideration
are required to transform
under the adjoint action of $so(d-1,2)\,$, which in physical terms means that the corresponding higher-spin gauge fields couple
to gravity in the way explained below (\ref{hslincur1}).

\vspace*{.3cm}

Leaving aside all the possible constrains
that will be imposed on the
cubic vertices when investigating gauge invariance of the action $S=S_0+g\,S_1+g^ 2\,S_2$ at order ${\cal O}(g^2)\,$, let us find all the possible independent ${\mathfrak{sp}}(2)$-invariant
contractions of  ${\mathfrak{sp}}(2)$ singlets given by two-row
rectangular diagrams of some particular lengths $\check{n}$ and $\check{m}$ with
$\check{n}\leqslant \check{m}\,$.
As it was explained above,
in the general case (of arbitrarily long Young
diagrams $f^{\check{n}}(Y)$ and $g^{\check{m}}(Z)$
with degree of homogeneity in $Y^A_{\alpha}$ and $Z^A_{\alpha}$ being $2{\check{n}}$ and $2{\check{m}}\,$, respectively),
all the independent polynomials in $s_{YZ}$
and $p_{YZ}$ produce independent
contractions. On the other hand, it is obvious
that finite Young diagrams cannot be contracted in
an infinite number of independent ways.
Moreover, it is clear that contractions with
sufficiently large powers of \footnote{In the following we will often
use the notation $s$ and $p$ in place of
$s_{YZ}$ and $p_{YZ}$ when no confusion can
arise.} $s$ and $p$ annihilate any given
Young diagrams, each being a monomial
of finite degree in $Y$ or $Z\,$,
like $f^{\check{n}}(Y)$ and $g^{\check{m}}(Z)\,$.
So, our goal is to study the independent contractions
for finite Young diagrams.
This problem can be solved by representation
theory methods, where
it amounts to taking tensor product of two
representations associated with $W^n$ and $W^m$
and decomposing the result into irreducible
two-row Young tableaux parts.
This being said, we will make a more direct
analysis that gives an explicit realization
of all the independent contractions in terms
of polynomials in the operators $s$ and $p\,$.

Given two $\mathfrak{sp}(2)$-singlets
$f^{\check{n}}(Y)$ and $g^{\check{m}}(Z)\,$
of degree $\check{n}$ and $\check{m}$ in
$Y$ and $Z$ respectively,
first note that the action
of a single $p_{YZ}$ operator on $f(Y)g(Z)$
contracts twice as many indices as $s_{YZ}$ does,
see (\ref{sandt}).
The total number of contracted indices
in one of the two Young tableaux will be
called the degree of contraction and denoted
by $k\,$. So, for
 the contraction $p^{\alpha}s^{\beta}\,$,
the degree of contraction is $k=2\alpha+\beta\,$.
Obviously, only contractions of the same degree
may be linearly dependent.
The next thing to note is that the significant
difference between $s^2$ and $p$ is that
$p$ contracts the same number of indices in the
first and in the second row of, say, the  first
Young tableau. In contrast to $p\,$,
$s^2$ contains terms that contract two indices
in only the first or the second row of $f^{\check{n}}(Y)\,$.

In general, let us consider the operator
${\cal O}_1^{(\alpha,\beta)}=p^{\alpha}s^{\beta}$.
The maximal number of indices  it contracts in
the first row of $f^{\check{n}}(Y)$ is
$M({\cal O}_1^{(\alpha,\beta)})=\alpha+\beta\,$.
Let us note that $M({\cal O})$, {\emph{i.e.}}
the number of indices contracted
in the first row of $f^{\check{n}}(Y)$
by an operator ${\cal O}\,$,
is a quantity that cannot
be changed by using Young symmetry properties
of $f^{\check{n}}(Y)\,$.
Now we consider the operator
${\cal O}_2^{(\alpha-1,\beta+2)}=p^{\alpha-1}s^{\beta+2}$
of the same degree as ${\cal O}_1^{(\alpha,\beta)}\,$.
The maximal number $M_2$ of indices contracted in the first row now is $\alpha+\beta+1\,$.
{}From the fact that ${\cal O}_2^{(\alpha-1,\beta+2)}$ contains the terms where $\alpha +\beta+1$
indices are contracted in the first row and
${\cal O}_1^{(\alpha,\beta)}$ does not,
it follows that  ${\cal O}_1^{(\alpha,\beta)}$ and
${\cal O}_2^{(\alpha-1,\beta+2)}$ produce
linearly independent contractions.
Following this logic, one can show that all the
contractions
$p^{\alpha}s^{\beta}$ of fixed degree
$k=2\alpha+\beta $ are independent when
$k\leqslant \check{n}\,$.
We call this {\em Case I} in what follows.

On the other hand,
in {\em Case II} when $k > \check{n}\,$,
the above logic is not applicable
because some operators ${\cal O}$ such as $s^k$
are such that $M({\cal O}) > \check{n}\,$,
namely they have the maximal number of contractions
in the first row of $f^{\check{n}}(Y)$ exceeding the
total number of indices available.
Still, the operators with $M({\cal O})\leqslant \check{n}$ are linearly independent,
following the logic explained above.
More precisely, all the operators
$p^{\alpha}s^{\beta}$
of fixed degree $k=2\alpha+\beta$ and having
$M=\alpha+\beta$ are independent for
$\alpha+\beta\leqslant \check{n}\,$.
Let us call them
{\em definitely-independent Case II}
operators.
What is less easy to see is that all the
remaining operators (\emph{i.e.} those that have
$M>\alpha+\beta$) of the same degree can be given
as linear combinations of those having
$M=\alpha+\beta \leqslant \check{n}\,$.
One can prove this proposition
from the associativity
of the tensor product, as follows.

Let us consider the operators belonging to
\emph{Case I} and let us compute how many of them
can contract two Young tableaux of respective
lengths $\check{n}$ and $\check{m}$
(with $\check{n}\leqslant \check{m}$)
and produce a resulting
Young tableau of length $\check{\ell}\,$.
The first obvious relation is
\begin{equation}
\label{deg}
\check{n}+\check{m}-k=\check{\ell}\; .
\end{equation}
Now we fix $\check{n}$, $\check{m}$,
$\check{\ell}\,$, and consequently $k\,$.
The independent contractions
belonging to \emph{Case I}
(so that $k\leqslant \check{n}$) are such that
$\check{m}\leqslant \check{\ell}\,$.
So, the operators in \emph{Case I} can be alternatively be specified by
 \begin{equation}
 \label{caseI}
 \check{n}\leqslant \check{m}\leqslant \check{\ell}
 \;.
 \end{equation}
By definition of {\emph{Case I}},
all the operators $p^{\alpha}s^{\beta}$ in this
case are independent, so the total number of
independent operators equals
to the number of partitions of $k$ as
$k=2\alpha+\beta$ with
non-negative $\alpha$ and $\beta\,$.
It is easy
to see that the number of such partitions is
\begin{equation}
\label{multiplicity}
N_I=\Big[\frac{k}{2}\Big]+1=\Big[\frac{\check{n}+\check{m}-\check{\ell}}{2}\Big]+1.
\end{equation}
This gives the desired multiplicity of contractions
of $f^{\check{n}}$ and $g^{\check{m}}$
that produce a Young tableau $h^{\check{\ell}}$
with $\check{\ell}\geqslant  \check{m}\,$.

Obviously, once the  multiplicities
(\ref{multiplicity}) for \emph{Case I},
(\ref{caseI}), are known,
and from the associativity of the tensor product,
one can derive the multiplicities of the
contractions of $f^{\check{n}}$ and $g^{\check{m}}$
that give rise to $h^{\check{\ell}}$ with
$\check{\ell} < \check{m}\,$.
This is nothing but \emph{Case II}
since $\check{\ell} < \check{m}\,$
is equivalent to $\check{k}>\check{n}$,
cf. (\ref{deg}).
{}From (\ref{multiplicity}) one can find that the multiplicity
in this case is
\begin{equation}
\label{multiplicity1}
N_{II}=\Big[\frac{\check{n}+\check{\ell}-\check{m}}{2}\Big]+1.
\end{equation}

Now we want to show that this multiplicity is
the number of definitely-independent operators
as was explained above, which will therefore
prove that the remaining operators are just linear
combinations of the
definitely-independent ones, thereby proving our
proposition.

So, we compute the number of definitely-independent
operators $p^{\alpha}s^{\beta}$ with
the fixed degree $k=2\alpha+\beta$ in
\emph{Case II} and having
$\alpha+\beta\leqslant \check{n}\,$.
This multiplicity is the
number of partitions of $k$ in the form
$k=2\alpha+\beta$ such that
$\alpha+\beta\leqslant \check{n}$ and where
both $\alpha$ and $\beta$ are non-negative
integers. It is not hard to show that this gives
exactly
\begin{equation*}
N=\check{n}-\Big[\frac{k}{2}\Big]=
\Big[\check{n}-\frac{k}{2}\Big]+1=\Big[\frac{\check{n}+\check{\ell}-\check{m}}{2}\Big]+1,
\end{equation*}
as anticipated.
To conclude, the definitely-independent
contractions indeed provide a basis of operators in
\emph{Case II}.

To summarize,
both possibilities $\check{m}\leqslant \check{\ell}$
and $\check{m}>\check{\ell}$
have been considered, and the bases of all
the possible contractions have been given.

\section{Trace-associativity or invariant-normed algebra condition}
\label{tac}

With the basis for independent contractions known,
we can find the solution to Fradkin-Vasiliev condition (\ref{FVcond3}) derived in
Section  \ref{FV}.
The on-shell  curvatures are $V$-transversal
because of (\ref{fot}) and (\ref{vtrans}),
thereby only the last term in the sum
(\ref{linaction}) of the  quadratic action remains non-zero on-shell,
 \begin{equation}
 \label{linaction3}
 S_0\approx \frac{1}{2}\int d^d\!x\, a(k,k-2)G_{MNPQ}
 R^{MB(k-2),ND(k-2)}R^P{}_{B(k-2),}{}^{Q}{}_{D(k-2)}\;,
 \end{equation}
and therefore nontrivial cubic interactions
are obtained by substituting
$R=R_1+R_2$ in the above expression,
instead of using the full action (\ref{linaction}).

The first on-mass-shell theorem can be rewritten as
\begin{equation*}
R_0^{A(k-1),B(k-1)}\approx E_0^ME_0^N C^{A(k-1),B(k-1);}{}_{MN}
\quad \text{or} \quad R_0^{\{k\}}\approx  E_0^ME_0^NC^{\{k\};}{}_{MN}.
\end{equation*}
Here $C^{\{k\};}{}_{MN}$ has two groups of indices: (i) $2(k-1)$ indices
having a symmetry of two-row rectangular Young
tableau denoted implicitly by $\{k\}$,
and (ii) two indices $M$ and $N$ antisymmetrized
by contraction with two frame fields.
It is important to split the indices of $C$ into
two groups, because only the indices $\{k\}$ are
sensitive to gauge transformations
-- see also \cite{Boulanger:2011se}.
Indeed, from
\begin{equation*}
\delta_1^{YM} R^{\{k\}}=-\xi^{\{n\}}\diamond R^{\{m\}}+R^{\{n\}}\diamond \xi^{\{m\}}
+{\cal O}(W^2)\;,
\end{equation*}
it follows that
\begin{equation}
\label{gtrw}
\delta_1^{YM} (E_0^{M}E_0^N C^{\{k\};}{}_{MN})=E_0^ME_0^N(-\xi^{\{n\}}\diamond C^{\{m\};}{}_{MN}
+C^{\{n\};}{}_{MN}\diamond \xi^{\{m\}}).
\end{equation}
Therefore
 \begin{equation}
 \label{linaction4}
\delta_1^{YM} S_0+\delta_0S_1\approx 2\int d^d\!x\, a(k,k-2)G_{MNPQ}
 R^{MB(k-2),ND(k-2)}\delta_1^{YM} R^P{}_{B(k-2),}{}^{Q}{}_{D(k-2)}
 +{\cal O}(W^3\xi)
 \end{equation}
can be rewritten as
\begin{eqnarray}
\delta_1^{YM} S_0+\delta_0S_1 &\approx & \int d^d\!x\, a(k,k-2)G_{MNPQ}E^RE^S
 C_{RS;}{}^{MN;B(k-2),D(k-2)}\times
\nonumber \\
&& E_0^TE_0^U(-\xi^{\{n\}}\diamond C^{\{m\};}{}_{TU}
+C^{\{n\};}{}_{TU}\diamond \xi^{\{m\}})^{PQ;}{}_{B(k-2),D(k-2)}+{\cal O}(W^3\xi)\;.
\label{cubictr1}
\end{eqnarray}
One can write  Lorentz indices everywhere
instead of (anti)-de Sitter indices
because of (i) the on-mass-shell theorem
states that curvatures are $V$-transversal on-shell;
and (ii) since the symbol $G_{MNPQ}$ defined in
(\ref{notation1})
contains an explicit contraction of the totally
antisymmetric $so(d-1,2)$ tensor with a
compensator $V\,$, so that all the remaining
indices of the antisymmetric tensor run only over
the $V$-transversal, or Lorentz, directions.
Using the identity \cite{Vasiliev:2001wa}
\begin{equation*}
E_0^C G_{A_1...A_k }=\frac1{(d-k+1)}\sum_{i=1}^{i=k}(-)^{i+k}\delta^C_{A_i}G_{A_1...\hat{A_i}...A_k}
\end{equation*}
one can show that
\begin{equation}
\label{id1}
G_{MNPQ}E_0^RE_0^SE_0^TE_0^U\propto \delta^{[RSTU]}_{[MNPQ]}G\;.
\end{equation}
We do not specify the precise coefficient
because it only depends on the dimension of
the space
and cancels in the following computations.

Let us focus on the first term in the bracket of (\ref{cubictr1}).
According
to (\ref{id1}) it can be rewritten as
\begin{equation}
\label{cubictr2}
\int d^dx \,G\,a(k,k-2)\delta^{[RSTU]}_{[MNPQ]}
 C_{RS;}{}^{MN;B(k-2),D(k-2)}
(-\xi^{\{n\}}\diamond C^{\{m\};}{}_{TU})^{PQ;}{}_{B(k-2),D(k-2)}\;.
\end{equation}
Due to the tracelessness of the Weyl tensor,
the indices $R$ and $S$ can be
contracted with $P$ and $Q$ only,
so we can rewrite (\ref{cubictr2}) as
\begin{equation}
\label{cubictr3}
\int d^dx \,G\, a(k,k-2)
 C^{MN;B(k-1),D(k-1)}
(-\xi^{\{n\}}\diamond C^{\{m\};}{}_{MN})_{B(k-1),D(k-1)}\;.
\end{equation}
Regrouping terms as in (\ref{FVcond3}) one finds that the Fradkin-Vasiliev
condition is equivalent to
\begin{eqnarray}
& a(k,k-2)
 C^{MN;B(k-1),D(k-1)}
(\xi^{\{n\}}\diamond C^{\{m\};}{}_{MN})_{B(k-1),D(k-1)}=\hspace*{4cm} &
\nonumber \\
&a(m,m-2)
 C^{MN;B(m-1),D(m-1)}
(C^{\{k\};}{}_{MN}
\diamond\xi^{\{n\}})_{B(m-1),D(m-1)}\;.&
\label{cubictr4}
\end{eqnarray}

Suppose that the particular $\diamond$-product
between $\xi^{\{n\}}$ and $C^{\{m\};}{}_{MN}$ is
realized as $k^k_{n,m}p^{\alpha}s^{\beta}$ and
produces a Young tableau that belongs to the
spin-$k$ sector, which implies
\begin{equation}
\label{ident1}
\check k=\check n+\check m-2\alpha-\beta\;.
\end{equation}
In the appendix it is shown in (\ref{ibp1})
that
\begin{align}
\notag
&C^{MN;B(k-1),D(k-1)}
(\xi^{\{n\}}p^{\alpha}s^{\beta} C^{\{m\};}{}_{MN})_{B(k-1),D(k-1)}=\\
\label{apref}
&\tfrac{(\alpha+\beta+1)}{(\alpha'+\beta +1)}\;
C^{MN;B(m-1),D(m-1)}
(C^{\{k\};}{}_{MN}\,p^{\alpha'}s^{\beta}\,\xi^{\{n\}})_{B(m-1),D(m-1)}\;,
\end{align}
where  $\alpha'=\check n-\alpha-\beta$.
Therefore, in order to solve (\ref{cubictr4})
the $\diamond$-product between $C^{\{k\}}$ and
$\xi^{\{n\}}$ should
have the form $k^m_{k,n}p^{\alpha'}s^{\beta}$ and
\begin{equation*}
a(k,k-2)\,k^{k}_{n,m}=\tfrac{(\alpha+\beta+1)}{(\alpha'+\beta+1)}\;a(m,m-2)\,k^{m}_{k,n}\;.
\end{equation*}
In terms  of the spins $m$, $n$, $k$ and the free parameter $\beta\,$,  it gives
\begin{equation}
\label{finalFV}
a(k,k-2)k^{k}_{n,m}=\tfrac{(n+m-k+\beta+1)}{(n+k-m+\beta+1)}\; a(m,m-2)\,k^{m}_{k,n}\;.
\end{equation}
This equation explicitly displays the implication of Fradkin-Vasiliev condition on the
free coefficients $a$ and $k$.
Obviously, if the particular $\diamond$ contraction
between $\xi^{\{n\}}$
and $C^{\{m\};}{}_{MN}$ is realized as $k^k_{n,m}p^{\alpha}s^{\beta}\,$, then
one can always find $k^{m}_{k,n}$ so as to satisfy
(\ref{finalFV}),
which means that every such contraction can be
promoted to a consistent higher-spin cubic
vertex. Having classified all the independent
contractions in Section \ref{star}, we
thereby classified all the independent higher-spin
cubic vertices.
\vspace*{.2cm}

Finally, it is easy to see that the number of independent
contractions given in Section \ref{star} coincides with the
number of possible non-abelian algebra deformations obtained in
\cite{Bekaert:2010hp} \footnote{We recall that, for a triplet of spins with
$s\leqslant s' \leqslant s''\,$, the non-abelian deformations of the
gauge algebra can give rise to vertices with a number of derivatives $k$
ranging from $k^{min}= s''+s'-s$ to $k_o^{max}= 2 s'-1$
for odd $\mathbf{s}:= s+s'+s''$ or to $k_e^{max}= 2 s'-2$
for even $\mathbf{s}\,$. Therefore the multiplicity of non-abelian vertices
is
$N_o = \frac{s+s'-s''+1}{2}\,\equiv \left[ \frac{s+s'-s''}{2}\,\right]+1$
for odd $\mathbf{s}$ and
$N_e = \frac{s+s'-s''}{2}\,$
for even $\mathbf{s}\,$, which exactly matches the multiplicity formula
found in Section \ref{star}.},
thereby proving that our list of independent non-abelian vertices in $AdS_d$
is exhaustive.

%**********************************
\section{(Mixed)-symmetry (partially)-massless fields}\label{sec:gravitational}
%*********************************************
In this section we discuss how to construct gravitational interactions in anti-de Sitter space for
gauge fields of various types\footnote{For some results on interactions of mixed-symmetry fields on flat background see \cite{Bekaert:2002uh,Boulanger:2004rx,Bekaert:2004dz,Metsaev:2005ar,Metsaev:2007rn}, as for anti-de Sitter space a few results are available \cite{Alkalaev:2010af,Zinoviev:2011fv,Boulanger:2011qt,Boulanger:2011se}.
Interactions of partially-massless fields has been studied recently in \cite{Deser:2012qg}
} and discuss briefly general non-abelian interactions. The simplest example is provided by a spin-$s$ partially-massless field of
depth-$t$. Partially-massless fields \cite{Deser:2001us} have the following higher-derivative
transformation law
\be\delta \phi_{\mu_1...\mu_s}=D_{\mu_1}...D_{\mu_t}\xi_{\mu_{t+1}...\mu_s}+\ldots \;,
\ee
where the parameter $t\in \{ 1,\ldots,s\} $
is called the depth and $...$ stands for the terms with less derivatives.
As shown in \cite{Skvortsov:2006at} a spin-$s$ partially-massless field of depth-$t$ can be described by a one-form connection that takes values in the irreducible
tensor representation of $so(d-1,2)$  defined by a two-row Young diagram
\be\delta W^{A(s-1),B(s-t)}=D_0 \xi^{A(s-1),B(s-t)}\,,\qquad\qquad \parbox{60pt}{\boldpic{\RectBRow{6}{4}{$s-1$}{$s-t$}}}\,.\ee
Massless fields arise at $t=1$. The equations of motion are similar to (\ref{fot})
\be R^{A(s-1),B(s-t)}=D_0W^{A(s-1),B(s-t)}\,,\qquad\qquad R^{A(s-1),B(s-t)}=E^M_0 E^N_0 C^{A(s-1)}{}_M{}^{,B(s-t)}{}_{N}\,,\ee
where the Weyl tensor for partially-massless field has the symmetry of \parbox{62pt}{\boldpic{\RectBRow{6}{5}{$s$}{$s-t+1$}}} and it is $V$-transverse.

As before we write the most general quadratic corrections to the field strength of the spin-$2$ field $W^{U,U}$
and to that of the partially-massless field $W^{A(s-1),B(s-t)}$
\begin{align*}
R^{U,U}&=\DO W^{U,U}+g_1 W^{A(s-2)U,B(s-t)}\wedge W\fdud{A(s-2)}{U}{,B(s-t)}\nonumber
+\\&\phantom{=\DO W^{U,U}+}+g_2 W^{A(s-1),B(s-t-1)U}\wedge W\fdu{A(s-1),B(s-t-1)}{U},\\
R^{A(s-1),B(s-t)}&=\DO W^{A(s-1),B(s-t)}+ W\fud{A,}{M}\wedge W^{MA(s-2),B(s-t)}+ W\fud{B,}{M}\wedge W^{A(s-1),MB(s-t-1)}\,.
\end{align*}
Note that there are two independent contributions to $R^{U,U}$. The quadratic correction to $R^{A(s-1),B(s-t)}$ is
just an $so(d-1,2)$-covariant derivative.
The quadratic actions for the graviton and partially-massless field read
\begin{align*}
S^{\{2\}}&=\alpha_2\int R^{U,U}\wedge R^{V,V}G_{UUVV}\\
S^{\{pm\}}&=\sum \alpha^{s,t}_{q,m}\int R^{UA(s-m-2)C(m),UC(q)B(s-q-2)}\wedge R\fudud{V}{A(s-m-2)}{C(m),VC(q)}{B(s-q-2)}V_{2q+2m}G_{UUVV}
\end{align*}
where $a^{s,t}_{q,m}$ are certain coefficients fixed up to an overall factor \cite{Skvortsov:2006at}, which we identify with $\alpha^{s,t}_{0,0}$.

Using the general formulae (\ref{YMgauge}), (\ref{curvtr}) and (\ref{FVcond3}),
one requires the gauge invariance of
the cubic terms on the free mass-shell, resulting in the condition
$\delta S^{\{2\}} + \delta S^{\{pm\}} =0 \,$, where
\begin{align*}
\delta S^{\{2\}}&=4g_1\alpha_2\overbrace{\int C_{uu,vv}\,\xi\fdud{a(s-2)}{v,}{b(t)} C^{a(s-2)uv,b(t)u}}^{\boldsymbol{A}}+
4g_2\alpha_2\overbrace{\int C_{uu,vv}\,\xi\fdu{a(s-1),b(t-1)}{v} C^{a(s-1)u,b(t-1)uv}}^{\boldsymbol{B}}\;,
\\
\delta S^{\{pm\}}&=\alpha^{s,t}_{0,0}(-2s)\boldsymbol{A}+\alpha^{s,t}_{0,0}(-\frac{2st}{s-1})\boldsymbol{B}
\;.
\end{align*}
Obviously, the condition
$\delta S^{\{2\}}+\delta S^{\{pm\}}=0$ admits a unique solution.
The ratio $g_1/g_2$ is a fixed number.
Therefore the freedom in $g_1$, $g_2$ does not lead to two different types of gravitational interactions.

Let us now comment of the general case of gravitational interactions of mixed-symmetry and/or partially-massless fields described by
one-form connections $W^{\Yy}$ with values in any irreducible tensor representation of $so(d-1,2)$
specified by a Young diagram $\Yy$ with rows of lengths $s_1,s_2,...,s_n$, $\Yy=\Y{s_1,...,s_n}\,$.
The dictionary between $W^{\Yy}$ and the metric-like formalism was given in
\cite{Alkalaev:2003qv, Alkalaev:2003hc, Boulanger:2008kw, Skvortsov:2009nv}.
The case of one-forms $W^{\Yy}$ does not cover the variety of all possible types of mixed-symmetry and
partially-massless fields. In order to take into consideration all gauge fields possible one has to include
gauge connections $W^{\Yy}$ that are forms of higher degree too. However, only one-form connections $W^{\Yy}$ can
give rise to a Lie algebra and only one-forms can source gravity
in the Fradkin-Vasiliev framework as in this case one can write $W^{\Yy}\wedge
W^{\Yy}$ contribution to the spin-$2$ field strength $R^{U,U}$ as we did above.
The most general ansatz reads
\begin{align}
R^{U,U}&=\DO W^{U,U}+\sum_{\displaystyle\Yy/\smallpic{\YoungA}} g_i W^{;U}\wedge W^{;U}\,,\label{YcruA}\\
R^{\Yy}&=\DO W^{\Yy}+ \sum_i W\fud{B,}{M}\wedge W^{A(s_1),...,MB(s_i-1),...} \,,\label{YcruB}
\end{align}
where in the first line the sum is over all possible ways to cut one cell from $\Yy$ such that the result is a
valid Young diagram. The number of such ways is equal to the number of blocks of $\Yy$.
If there are no rows in $\Yy$ that have equal length, then the sum is over all rows and in the $i$-th summand
one isolates one index in the $i$-th row, denotes it by $U$ and contracts the rest of the indices pairwise. The
deformation of $R^{\Yy}$ is just a covariant derivative with respect to dynamical spin-$2$ connection
$W^{U,U}$.

The linear equations of motion for $W^\Yy$ read
\cite{Alkalaev:2003qv, Boulanger:2008up, Boulanger:2008kw,Skvortsov:2009zu, Skvortsov:2009nv}
\be
R^\Yy=E^M_0 E^N_0 \Pi_{MN}(C^\Xx)^{\Yy}\,,
\ee
where the generalized Weyl tensor $C^\Xx$ is an irreducible $so(d-1,2)$-tensor
having the symmetry of $\Xx=\Y{s_1+1,s_2+1,s_3,...,s_n}$ and the projector $\Pi_{MN}$ isolates
two indices of $C$ and projects onto $\Yy$. The Weyl tensor for generic mixed-symmetry field
is not fully-transverse and satisfies more complicated $V$-dependent constraints,
\cite{Alkalaev:2003qv, Boulanger:2008up, Boulanger:2008kw, Skvortsov:2009zu, Skvortsov:2009nv},
which implies that $C$ contains more than one Lorentz component in general.
This is not the case for totally-symmetric (partially)-massless fields.

In order for the gravitational interactions of $W^\Yy$ to exist in the first nontrivial order one
has to prove that there is enough free coefficients to impose the invariance of the cubic vertex on the
free mass-shell, (\ref{cubicinv}). We will give an argument that this is indeed true despite the fact that
the quadratic actions are not known in full generality. In general to construct a Lagrangian the
connection $W^\Yy$ has to be supplemented with certain additional fields,
see e.g. \cite{Zinoviev:2008ve, Zinoviev:2009vy, Zinoviev:2009gh} for specific examples.
Fortunately, to
check the gauge invariance of the cubic vertex we only need to know
the on-shell action, \emph{i.e.} the terms
 in the action to which the generalized Weyl tensor contribute,
\be
\label{Yactiononshell}
\left.(S^{\{2\}}+S^{ \{\Yy\} })\right|_{on-shell}=\alpha_2\int R^{U,U}\wedge
R^{V,V}G_{UUVV}+\sum_{\displaystyle\Yy/\parbox{6pt}{\smallpic{\YoungAA}}} \alpha_n\int R^{;UU}
\wedge R^{;VV} G_{UUVV} \,,
\ee
where the sum is over all possible ways to isolate two anti-symmetric indices in tensor with the symmetry
of $\Yy$, these are to be contracted with $G_{UUVV}$, the rest are contracted pairwise.
These leading terms can be extracted from the results
of \cite{Alkalaev:2003qv, Alkalaev:2005kw, Alkalaev:2006rw}. That the Weyl tensor is not
fully $V$-transverse imposes severe restrictions on such terms. Indeed, one would naively add to
(\ref{Yactiononshell}) the terms where in addition to a pair of anti-symmetrized indices one isolates a
group of symmetric indices to be contracted with $V$. These additional $V$ contractions may be nonzero as
the Weyl tensor in not fully $V$-transverse. Taking then the variation of (\ref{Yactiononshell}), one finds
\be
\label{varYactiononshell}
\left.\delta (S^{\{2\}}+S^{\{\Yy\} })\right|_{on-shell}\sim\alpha_2\int [R,\xi]^{UU}\wedge
R^{V,V}G_{UUVV}+\sum_{\displaystyle\Yy/\parbox{6pt}{\smallpic{\YoungAA}}} \alpha_i\int [R,\xi]^{;UU} \wedge
R^{;VV} G_{UUVV} \,,
\ee
where $[R,\xi]$ can be read off from (\ref{YcruA})-(\ref{YcruB}) according to general formulae of
Section \ref{FV}.
One observes that $\xi^{A,B}$ contributes only to $\delta S^{\{\Yy\}}$  and not to $\delta S^{\{2\}}\,$.
Therefore, $\xi^{A,B}$-variation must vanish on its own. Indeed, that there is no $\parbox{6pt}{\smallpic{\YoungAA}}$ in the symmetric tensor product $Sym(\Xx\otimes\Xx)$ for any $\Xx$ implies that any singlet built
of $\xi^{A,B}$ and two Weyl tensors $C^\Xx$ is identically zero.
Now we have to cancel the $\xi^\Yy$-part of the variation. Note that $\delta S^{\{2\}}$ has no $V$ explicitly
besides in $G_{UUVV}$ since neither the deformation (\ref{YcruA}) nor the spin-$2$ action
contain $V\,$.
The latter implies that
$\delta S^{\{\Yy\}}$ and hence the on-shell part of $S^{\{\Yy\}}$ must not have any explicit $V$-contractions. This justifies the form of (\ref{Yactiononshell}).
Then, using the symmetric basis for Young diagrams it is easy to see that the sums in (\ref{YcruA}) and (\ref{YcruB}) produce pairwise identical
terms in $\delta S^{\{2\}}$ and $\delta S^{\{\Yy\}}\,$.
In particular all the terms in the sum of (\ref{Yactiononshell}) vanish except for the one where two anti-symmetrized indices $UU$ belong to the
first two rows of $\Yy$. Equivalently, using the freedom of adding total derivatives of the form $\int D_0(R R V G)$, \cite{Alkalaev:2003qv, Alkalaev:2003hc,Alkalaev:2005kw,Alkalaev:2006rw}, one can reduce the number of terms in the sum of (\ref{Yactiononshell}) to a single term described above. Again all the ratios $g_i/g_j$ are certain fixed numbers and hence the gravitational interactions are essentially unique.

Let us make some comments about general non-abelian interactions of (mixed)-symmetry and/or partially-massless fields. We restrict ourselves to those gauge fields in the metric-like approach that are described by one-form connections $W^\Yy$ within the frame-like approach. The condition for the variation to vanish amounts to 
\be
 (A|B\diamond C)- (A\diamond B | C)=0\,,
 \label{trass}
 \ee
 where $A$, $B$, $C$ correspond to two Weyl tensors and one gauge parameter; $\diamond$  stands for some particular way of contracting indices; $(x|y)$
  takes the singlet part. Given some $A\diamond B$ one can always adjust
  the contraction  $B \diamond C$ such that (\ref{trass}) is true.
   As we argued above, see also \cite{Boulanger:2011se}, already the gravitational interactions restrict the freedom of adding topological terms $\int D_0(R R V G)$ in such a way that the Weyl tensor has no $V$-contractions in the on-shell action. The appearance of the Weyl tensor contracted with a number of compensators $V$ would invalidate the arguments above. Therefore we see that each independent way of contracting indices among two connections $W^{\Yy_1}$ and $W^{\Yy_2}$ gives rise to a consistent cubic vertex, which
   is non-abelian by definition.
   
\paragraph*{Similitude with Yang--Mills and invariant-normed algebra.}
The parallel between the above discussion and the spin-1 case is obvious,
and we have seen that it is always possible to contract the indices of rectangular two-row
Young tableaux in such a way that the resulting cubic action is consistent at that order.
This becomes clear if one highlights the similitude of the Fradkin-Vasiliev construction
 with the Yang-Mills one. The Fradkin-Vasiliev procedure is precisely inspired by the
Yang-Mills, geometric treatment of gauge systems.
Consider, as a starting point, a positive sum of $n$ Maxwell's actions for a set of one-form gauge fields
$\{A^a\}_{a=1,\ldots,n}$
\begin{eqnarray}
S_0[A^a] &=& \int_{M_4} \langle F_1 , F_1 \rangle \;
\;\equiv~\int_{M_4} k_{ab}\; F_1^a\wedge * F_1^b \;, \quad
\quad F_1^a := {\rm d}A^a\;,
\end{eqnarray}
where $k_{ab}$ is diagonalized to $k_{ab} = c_a \delta_{ab}\,$ with $c_a >0$
for the sake of unitarity.
In order to introduce cubic interactions one performs the substitution
\begin{eqnarray}
F^a_1\quad \longrightarrow \quad F^a &:=& F^a_1 + g\,f^a{}_{bc}A^a A^b \;
\end{eqnarray}
inside $S_0$ while disregarding quartic terms, as we did with the Fradkin-Vasiliev
procedure. By definition of $F^a$ and because $A^a$ are one-forms, one has
\begin{eqnarray}
f^a{}_{bc} = -f^a{}_{cb}\;,
\end{eqnarray}
which defines an internal anti-commutative algebra ${\cal A}$ with basis
elements $\{e_a\}$ and product law $\diamond\;$ given by
\begin{eqnarray}
e_a\diamond e_b = f^c{}_{ab}\,e_c = -e_b\diamond e_c\;.
\end{eqnarray}
As is well-known and easy to see -- a cohomological derivation can be found in \cite{Barnich:1993pa},
the resulting deformed action $S_0+S_1$ is consistent
to order ${\cal O}(g)$ provided one has the following antisymmetry condition
\begin{eqnarray}
 f_{abc}:= k_{ad}\,f^d{}_{bc}=f_{[abc]}\;.
 \end{eqnarray}
In turn, this means that ${\cal A}$ is an invariant-normed (sometimes called graded-symmetric)
algebra, namely
\begin{eqnarray}
 \forall x,y,z \;\in \;{\cal A}\;, \quad ( x\diamond y, z ) =
( x\, ,\, y \diamond z ) \;,
\end{eqnarray}
where the norm is defined by
\begin{eqnarray}
 (x,y) = k_{ab}\,x^ay^b\;,\quad x=x^a\,e_a\;, \quad y=y^a\,e_a\;. 
 \end{eqnarray}
Given some constants $f^a{}_{bc}$ that satisfy $f^a{}_{bc} = -f^a{}_{cb}\,$,
it is always possible to find $f_{abc}$ that are completely antisymmetric,
thereby producing a consistent cubic vertex.

\vspace*{.2cm}

\noindent The story repeats itself in the higher-spin context where the internal index
$a$ is replaced with a rectangular two-row tensor representation of $so(d-1,2)\,$.
The fact that the Yang-Mills index $a$ now has an inner structure in the higher-spin case
implies that there is a multiplicity of choices for the $\diamond$-products
or equivalently for the constants $f^a{}_{bc}\,$'s
-- and where one may need to add a color index on every higher-spin gauge fields
in order to ensure the antisymmetry of $f^a{}_{bc}=-f^a{}_{cb}\,$; this is the case
for example when the $\diamond$-product is given by pure $p$ contractions
in the sector of odd spins. The determination of these multiplicities was done
in Section \ref{star} or could be obtained from group theory.

\vspace*{.2cm}

\noindent As in the spin-1 Yang-Mills case, the invariant-norm condition
$( x\diamond y, z ) = ( x\, ,\, y \diamond z )$ can also be achieved
in the higher-spin case, for every independent  choice of $\diamond$-product.

\vspace*{.2cm}

\noindent What will severely constrain the $\diamond$-product is the Jacobi condition
that arises at second order in the coupling constant $g\,$,
\begin{eqnarray}
 f^a{}_{b[c}f^b{}_{de]} = 0\;.
 \end{eqnarray}
In the spin-1 case, it implies that $f^a{}_{bc}$ define the structure constants
of a semi-simple Lie algebra.

%**********************************h
\section{Conclusions}\label{sec:conclusions}
%*********************************************

In this paper we have classified and explicitly built
all the possible non-abelian cubic vertices among totally
symmetric gauge fields in $AdS_d\,$.
The universal property of the universal enveloping algebra
guarantees that there exists
only one gauge algebra that can lead to an associative higher-spin
algebra, and that the latter precisely coincides with the
algebra used by Vasiliev in \cite{Vasiliev:2003ev}
for the construction of his nonlinear equations.
When pushing the analysis of vertices to the next, quartic order,
one typically finds that the internal algebra with
(graded)-antisymmetric structure constant should obey the Jacobi identity,
which is automatically satisfied if the commutator arises from the underlying
associative structure, see e.g. the discussion and the results
reviewed in \cite{Bekaert:2006us}. It is likely that
the only cubic vertex that has a chance to be promoted
to the next order is the one associated with the so-called
``$s$-contraction"
rule of Section \ref{star}, where the latter is the germ for
the associative algebra used in \cite{Vasiliev:2003ev} via
the Moyal-Weyl star-product formula (\ref{starproduct}). There
is still a loophole in that there can exist a higher-spin
algebra, which is essentially a Lie algebra. For example, a Poisson contraction
of the Vasiliev algebra, i.e. the one where $\exp{\hbar s}$, (\ref{starproduct}), is
expanded to the leading order in a formal non-commutativity parameter $\hbar$ would seem a good
candidate. However, the Poisson contraction is inconsistent even at the cubic level, as was pointed in \cite{Fradkin:1986qy} for the $4d$ case and the statement is valid for any $d$. The technique developed in this paper can be used to examine the question of uniqueness of higher-spin algebra in full generality and we leave it for a future publication.

We view the determination of cubic vertices
as one way to gain insight into the structure
and uniqueness of the full theory proposed in
\cite{Vasiliev:1990en,Vasiliev:1992av,Vasiliev:2003ev}.
In this sense, our results strongly confirm the
belief that Vasiliev's construction is the unique
way to obtain fully nonlinear and consistent interactions
among higher-spin gauge fields.
In the spirit of the Noether procedure for consistent interactions this implies that Vasiliev's theory can be viewed
as the gauging of the rigid star-product algebra $hu(1|2:[d-1,2])\,$,
and that this is the only way to construct a fully nonlinear
theory starting in perturbation around a fixed (here $AdS_d$)
background.

We showed that the (partially)-massless (mixed)-symmetry gauge fields that are described by one-form
connections $W^\Yy$ valued in irreducible representations of $so(d-1,2)$ can interact with gravity.
This gives a nontrivial indication that within the metric-like approach one will face certain difficulties in
trying to make interact with gravity those gauge fields
that are described by gauge connections of higher degree within the frame-like approach we use.
It seems that the frame-like approach contains more information about interactions even at the linear level.
Another example of this phenomenon was observed in \cite{Skvortsov:2010nh}, where a simple argument prevents
constructing Lagrangians for certain types of fermionic fields, which is highly nontrivial to see in the
metric-like approach \cite{Buchbinder:2009pa, Zinoviev:2009wh}.
The gravitational interactions for fields that are described by forms of higher degree
in the frame-like approach are severely constrained, see e.g. \cite{Bekaert:2000ba}
and references therein.
The gauge transformations for the $p$-form gauge fields can only be
deformed \`a la Chaplin-Manton \cite{Chapline:1982ww} or
Freedman-Townsend \cite{Freedman:1980us},
so that the gauge algebra in the $p$-form sector remains abelian although the
gauge transformations are modified non-trivially, sometimes even non-polynomially.

We also argued that those mixed-symmetry and/or partially-massless fields that are described by one-form
connections within the frame-like approach can have nonabelian interactions among themselves and again the number of nonabelian vertices should be given by tensor product multiplicities. Within the metric-like approach such gauge fields have the gauge parameter whose Young diagram is obtained by removing cells from the first row of the Young diagram of the field potential. For the rest of gauge fields, which are all nonunitary in AdS, \cite{Metsaev:1997nj,Brink:2000ag}, within the metric-like approach one still can write a lot of terms for the most general ansatz for the cubic vertex,
but we expect that the gauge invariance will result in a trivial solution only.

The technique used in the paper can be generalized to various cases of (partially)-massless fields
\cite{Skvortsov:2006at} and (mixed)-symmetry fields
\cite{Alkalaev:2003qv,Alkalaev:2006rw,Boulanger:2008kw,Skvortsov:2009nv,Skvortsov:2009nv}.

\vspace*{1cm}
\noindent \textbf{Note added}

During the final stage when the file was being prepared for submission to the arxives,
the paper \cite{Joung:2012hz} appeared where cubic vertices
for (partially-)massless fields are constructed, following a different procedure.
The tools presented there allow the construction of all possible types of vertices.
The nature of the gauge algebras associated with the vertices is not clear, though,
except in the Born--Infeld cases for obvious reasons. After identifying which of the vertices in \cite{Joung:2012hz} are non-abelian, it would be interesting to see if their number 
 is
indeed given by certain tensor product multiplicities as we showed in the present paper. Some simple examples show that this is the case.

\section*{Acknowledgements}

We thank Pierre Cartier and Per Sundell for useful discussions
related to universal enveloping algebra. We also thank 
Philippe Spindel and Konstadinos Siampos
for interesting discussions.
The work of N.B. and D.P. was supported in part by
an ARC contract No. AUWB-2010-10/15-UMONS-1. The work of E.S. was
supported  in part by the Alexander von Humboldt Foundation, by RFBR grant
No.11-02-00814, 12-02-31837 and Russian President grant No. 5638.

\section{Appendix}

Here we introduce some notations and
prove certain identities required
for solving the Fradkin-Vasiliev condition
for a general cubic vertex.

Let us first introduce a projection $f$ operation  which antysimmetrizes two indices that belong to different
rows of a Young diagram
\begin{equation*}
f(W)=W^{M_1N_1;A(m-1),B(m-1)}=
\frac{1}{2}\left(W^{A(m-1)M_1,B(m-1)N_1}-
W^{A(m-1)N_1,B(m-1)M_1}\right).
\end{equation*}
This operation is relevant to $p$ contraction
\begin{align*}
 &W^np W^m= W^{A(n-1)M,B(n-1)N}\left(W^{A(m-1)}{}_{M}{}^{,B(m-1)}{}_{N}-W^{A(m-1)}{}_{N}{}^{,B(m-1)}{}_{M}\right)=\\
&=2W^{A(n-1)M,B(n-1)N}W_{MN}{}^{;A(m-1),B(m-1)}=2W^{MN;A(n-1),B(n-1)}W_{MN}{}^{;A(m-1),B(m-1)}.
\end{align*}

An iterative application of an $f$-projector $\alpha$ times
gives
\begin{equation}
\label{falpha}
f^{\alpha}(W^{A(m),B(m)})=W^{M_1N_1,M_2N_2,\dots
M_{\alpha}N_{\alpha};A(\gamma),B(\gamma)}.
\end{equation}
where $\gamma=m-\alpha$.
It is straightforward to check that the right hand side of
(\ref{falpha}) possesses symmetry of $\mathbf{Y}(\alpha,\alpha)$
 in the antisymmetric basis  in the first
group of $2\alpha$ indices in the same time
having symmetry of $\mathbf{Y}(\gamma,\gamma)$
in the symmetric basis in the remaining indices.

By  iterative application of
\begin{equation*}
W^{M_1N_1,\dots
M_{\alpha-1}N_{\alpha-1},AB;A(\gamma),B(\gamma)}=
\frac{1}{2}\cdot\frac{\gamma+2}{\gamma+1}\cdot
W^{M_1N_1,\dots
M_{\alpha-1}N_{\alpha-1};A(\gamma+1),B(\gamma+1)}
\end{equation*}
one can find
\begin{equation}
\label{symm1}
W^{\overbrace{\scriptstyle AB,\dots
AB,AB}^{\alpha};A(\gamma),B(\gamma)}=
\frac{1}{2^{\alpha}}\cdot\frac{\alpha+\gamma+1}{\gamma+1}
\cdot
W^{A(\alpha+\gamma),B(\alpha+\gamma)}.
\end{equation}

Another useful representation appears when one symmetrizes
only $M$ and $N$ indices among each other in (\ref{falpha}) resulting in
\begin{equation}
\label{symm2}
W^{M_1N_1,M_2N_2,\dots
M_{\alpha}N_{\alpha};A(\gamma),B(\gamma)}
\rightarrow
W^{\overbrace{\scriptstyle MN,MN,\dots
MN}^{\alpha};A(\gamma),B(\gamma)}.
\end{equation}
This tensor has a symmetry of two row rectangular Young diagram
in symmetric convention in both groups of indices. Obviously
the same symmetry can be reached in a different
way
\begin{equation}
\label{symm3}
f^{\gamma}(W^{M(\alpha+\gamma),N(\alpha+\gamma)})=W^{A_1B_1,A_2B_2,\dots
A_{\gamma}B_{\gamma};M(\alpha),N(\alpha)}
\rightarrow
W^{\overbrace{\scriptstyle AB,AB,\dots
AB}^{\gamma};M(\alpha),N(\alpha)},
\end{equation}
which implies that right hand sides of (\ref{symm2}) and (\ref{symm3}) are proportional,
that is
\begin{equation}
\label{symm4}
W^{\overbrace{\scriptstyle MN,MN,\dots
MN}^{\alpha};A(\gamma),B(\gamma)}
= X(\alpha, \gamma)
W^{\overbrace{\scriptstyle AB,AB,\dots
AB}^{\gamma};M(\alpha),N(\alpha)}
\end{equation}
with some $X(\alpha,\gamma)$.
To find $X$ we symmetrize $M$ with $A$ and $N$
with $B$ in both sides of (\ref{symm4}), which, according to
(\ref{symm1}) results in
\begin{equation}
\label{symm5}
\frac{1}{2^{\alpha}}\cdot\frac{\alpha+\gamma+1}{\gamma+1}
\cdot
W^{A(\alpha+\gamma),B(\alpha+\gamma)}=
X(\alpha,\gamma)
\frac{1}{2^{\gamma}}\cdot\frac{\alpha+\gamma+1}{\alpha+1}
\cdot
W^{A(\alpha+\gamma),B(\alpha+\gamma)},
\end{equation}
which in turn implies
\begin{equation}
\label{sym6}
X(\alpha,\gamma)=
2^{\gamma-\alpha}\frac{\alpha+1}{\gamma+1}.
\end{equation}
So, we introduce a notation
\begin{equation*}
W^{M(\alpha),N(\alpha);A(\gamma),B(\gamma)}=
W^{A(\gamma),B(\gamma);M(\alpha),N(\alpha)}=
\end{equation*}
\begin{equation}
\label{sym7}
\frac{2^{\alpha}}{\alpha+1}W^{\overbrace{\scriptstyle MN,MN,\dots
MN}^{\alpha};A(\gamma),B(\gamma)}=
\frac{2^{\gamma}}{\gamma+1}W^{\overbrace{\scriptstyle AB,AB,\dots
AB}^{\gamma};M(\alpha),N(\alpha)}.
\end{equation}
One can proceed in the same manner by breaking each small sub-Young diagram
into even smaller pieces using the same formulas.

\paragraph{Computation}
Our goal is to find out how to relate two terms of (\ref{cubictr4}).
Terms
\begin{equation}
\label{app1}
 C^{MN;B(k-1),D(k-1)}
(-\xi^{\{n\}}\diamond C^{\{m\};}{}_{MN})_{B(k-1),D(k-1)}
\end{equation}
and
\begin{equation}
\label{app2}
 C^{MN;B(m-1),D(m-1)}
(C^{\{k\};}{}_{MN}\diamond\xi^{\{n\}})_{B(m-1),D(m-1)}
\end{equation}
are proportional and our aim is to find the proportionality coefficient. The
$M$ and $N$ indices are not involved in $\diamond$-product. They are used just
to contract two Weyl tensors in the same way in both expressions. So
we can omit them and treat the Weyl  tensors as having effectively two indices less each.

As a warm up exercise let us find the proportionality coefficient for the case
when $\diamond$ is represented by the $p$ contraction only in some power $\alpha$.
Let us also introduce $\check k=k-1$, $\check n= n-1$ and $\check m=m-1$. In this terms
$2\alpha=\check n+\check m-\check k$.
\begin{align*}
& C^{MN;B(k-1),D(k-1)}
(\xi^{\{n\}}p^{\alpha} C^{\{m\};}{}_{MN})_{B(k-1),D(k-1)}\rightarrow\\
 &C^{B(k-1),D(k-1)}
(\xi^{\{n\}}p^{\alpha} C^{\{m\}})_{B(k-1),D(k-1)}=\\
&=
2^\alpha C_{A(\check k),B(\check k)}\xi^{CD,...,CD;A(\check n-\alpha),B(\check n-\alpha)}C_{CD,...,CD}{}^{;A(\check m-\alpha),B(\check m-\alpha)}=\\
&=2^\alpha C_{A(\check m-\alpha)U(\check n-\alpha),B(\check m-\alpha)V(\check n-\alpha)}\xi_{AB,...,AB}{}^{;U(\check n-\alpha),V(\check n-\alpha)}C^{A(\check m),B(\check m)}=\\
&=(\alpha+1)C_{A(\check m-\alpha)U(\check n-\alpha),B(\check m-\alpha)V(\check n-\alpha)}\xi_{A(\alpha),B(\alpha)}{}^{;U(\check n-\alpha),V(\check n-\alpha)}C^{A(\check m),B(\check m)}=\\
&=\frac{2^{\check n-\alpha+1}(\alpha+1)}{(\check n-\alpha+1)}
C_{A(\check m-\alpha)U(\check n-\alpha),B(\check m-\alpha)V(\check n-\alpha)}\xi_{A(\alpha),B(\alpha)}{}^{;UV,...,UV}C^{A(\check m),B(\check m)}=\\
&=\frac{2^{\check n-\alpha+1}(\alpha+1)}{(\check n-\alpha+1)}
C_{UV,...,UV;A(\check m-\alpha),B(\check m-\alpha)}\xi_{A(\alpha),B(\alpha)}{}^{;UV,...,UV}C^{A(\check m),B(\check m)}=\\
&=\frac{\alpha+1}{\check n-\alpha+1}(C^{\{k\}}p^{\check n-\alpha}\xi^{\{n\}})_{B(m-1),D(m-1)}C^{B(m-1),D(m-1)}
\rightarrow\\
& \frac{\alpha+1}{\alpha'+1}C^{MN;B(m-1),D(m-1)}
(C^{\{k\};}{}_{MN}p^{\alpha'}\xi^{\{n\}})_{B(m-1),D(m-1)},
\end{align*}
where $\alpha'=\check n-\alpha$.

Analogously one can show that
\begin{align}
\notag
&C^{MN;B(k-1),D(k-1)}
(\xi^{\{n\}}p^{\alpha}s^{\beta} C^{\{m\};}{}_{MN})_{B(k-1),D(k-1)}=\\
\label{ibp1}
&\frac{\alpha+\beta+1}{\alpha'+\beta +1}
C^{MN;B(m-1),D(m-1)}
(C^{\{k\};}{}_{MN}p^{\alpha'}s^{\beta}\xi^{\{n\}})_{B(m-1),D(m-1)},
\end{align}
where $\check k=\check n+\check m-2\alpha-\beta$ and $\alpha'=\check n-\alpha-\beta$.
To show this, let us note that $s^{\beta}$ is
\begin{equation*}
\xi s^{\beta}C=\xi(Y)\left(\frac{\partial^2}{\partial Y^A_1\partial Z_{2A}}-
\frac{\partial^2}{\partial Y^A_2\partial Z_{1A}}\right)^{\beta}C(Z)=
\end{equation*}
\begin{equation}
\label{sbeta}
\xi(Y)\sum_{i=0}^{\beta}\frac{(-)^i\beta!}{i!(\beta-i)!}
\left(\frac{\partial^2}{\partial Y^A_1\partial Z_{2A}}\right)^{\beta-i}
\left(\frac{\partial^2}{\partial Y^A_2\partial Z_{1A}}\right)^{i}C(Z).
\end{equation}

Each term of the expansion (\ref{sbeta}) has non-zero projection to the
space of tensors with a  symmetry
encoded by the rectangular Young diagram $\mathbf{Y}_{r}$ as well as other projections
encoded by non-recatangular Young diagrams $\mathbf{Y}_{nr}$.
Since in (\ref{ibp1}) $\xi s^{\beta}C$ appears only contracted
with other tensor valued in $\mathbf{Y}_{r}$,
each term of the expansion (\ref{sbeta})
contributes only with its $\mathbf{Y}_{r}$-shaped part.
This allows us to keep track of only the first term in (\ref{sbeta}),
while the others give some fixed proportional contributions. The following computation
relates the first term of the left hand side of (\ref{ibp1})  and
the last term of the right hand side of (\ref{ibp1})
\begin{equation*}
C^{MN;B(k-1),D(k-1)}
(\xi^{\{n\}}p^{\alpha}s^{\beta} C^{\{m\};}{}_{MN})_{B(k-1),D(k-1)}\Big|_{1st}\rightarrow
\end{equation*}
\begin{equation*}
2^{\alpha}C_{A(\check n+\check m-2\alpha-\beta),B(\check n+\check m-2\alpha-\beta)}
\xi^{CD\dots CD;A(\check n-\alpha-\beta)M(\beta),B(\check n-\alpha)}
C_{CD\dots CD;}{}^{A(\check m-\alpha),}{}^{B(\check m-\alpha-\beta)}{}_{M(\beta)}=
\end{equation*}
\begin{equation*}
2^{\alpha}C_{A(\check n+\check m-2\alpha-2\beta)K(\beta),B(\check n+\check m-2\alpha-2\beta)L(\beta)}
\times
\end{equation*}
\begin{equation*}
\xi^{CD\dots CD;A(\check n-\alpha-\beta)M(\beta),B(\check n-\alpha-\beta)L(\beta)}
C_{CD\dots CD;}{}^{A(\check m-\alpha-\beta)K(\beta),}{}^{B(\check m-\alpha-\beta)}{}_{M(\beta)}=
\end{equation*}
\begin{equation*}
(\alpha+1)C_{A(\check m-\alpha-\beta)U(\check n-\alpha-\beta)K(\beta),B(\check m-\alpha-\beta)V(\check n-\alpha-\beta)L(\beta)}
\times
\end{equation*}
\begin{equation*}
\xi_{A(\alpha),B(\alpha);}{}^{U(\check n-\alpha-\beta)M(\beta),V(\check n-\alpha-\beta)L(\beta)}
C^{A(\check m-\beta)K(\beta),}{}^{B(\check m-\beta)}{}_{M(\beta)}=
\end{equation*}
\begin{equation*}
(\alpha+1)C_{A(\check m-\alpha-\beta)U(\check n-\alpha-\beta)K(\beta),B(\check m-\alpha-\beta)V(\check n-\alpha-\beta)L(\beta)}
\times
\end{equation*}
\begin{equation*}
\frac{(\beta+1)(\check n-\alpha-\beta+1)}{(\check n-\alpha+1)}
\xi_{A(\alpha),B(\alpha);}{}^{M(\beta)L(\beta);U(\check n-\alpha-\beta),V(\check n-\alpha-\beta)}
C^{A(\check m-\beta)K(\beta),}{}^{B(\check m-\beta)}{}_{M(\beta)}=
\end{equation*}
\begin{equation*}
(-1)^{\alpha}
(\alpha+1)\frac{(\beta+1)(\check n-\alpha-\beta+1)}{(\check n-\alpha+1)}
C_{A(\check m-\alpha-\beta)U(\check n-\alpha-\beta)K(\beta),B(\check m-\alpha-\beta)V(\check n-\alpha-\beta)L(\beta)}
\times
\end{equation*}
\begin{equation*}
\xi_{B(\alpha),A(\alpha);}{}^{M(\beta)L(\beta);U(\check n-\alpha-\beta),V(\check n-\alpha-\beta)}
C^{A(\check m-\beta)K(\beta),}{}^{B(\check m-\beta)}{}_{M(\beta)}=
\end{equation*}
\begin{equation*}
(-1)^{\alpha}
(\alpha+1)\frac{(\beta+1)(\check n-\alpha-\beta+1)}{(\check n-\alpha+1)}
\frac{\alpha+\beta+1}{(\alpha+1)(\beta+1)}
C_{A(\check m-\alpha-\beta)U(\check n-\alpha-\beta)K(\beta),B(\check m-\alpha-\beta)V(\check n-\alpha-\beta)L(\beta)}
\times
\end{equation*}
\begin{equation*}
\xi_{B(\alpha)}{}^{M(\beta),}{}_{A(\alpha)}{}^{L(\beta);U(\check n-\alpha-\beta),V(\check n-\alpha-\beta)}
C^{A(\check m-\beta)K(\beta),}{}^{B(\check m-\beta)}{}_{M(\beta)}=
\end{equation*}
\begin{equation*}
(-1)^{\alpha}
\frac{(\check n-\alpha-\beta+1)(\alpha+\beta+1)}{(\check n-\alpha+1)}
\frac{2^{\check n-\alpha-\beta}}{(\check n-\alpha-\beta+1)}
C_{UV\dots UV;A(\check m-\alpha-\beta)K(\beta),B(\check m-\alpha-\beta)L(\beta)}
\times
\end{equation*}
\begin{equation*}
\xi_{B(\alpha)}{}^{M(\beta),}{}_{A(\alpha)}{}^{L(\beta);UV\dots UV}
C^{A(\check m-\beta)K(\beta),}{}^{B(\check m-\beta)}{}_{M(\beta)}=
\end{equation*}
\begin{equation*}
(-1)^{\alpha}(-1)^{\alpha+\beta}2^{\check n-\alpha-\beta}
\frac{\alpha+\beta+1}{\check n-\alpha+1}
C_{UV\dots UV;A(\check m-\alpha-\beta)K(\beta),B(\check m-\alpha-\beta)L(\beta)}
\times
\end{equation*}
\begin{equation*}
\xi_{A(\alpha)}{}^{L(\beta),}{}_{B(\alpha)}{}^{M(\beta);UV\dots UV}
C^{A(\check m-\beta)K(\beta),}{}^{B(\check m-\beta)}{}_{M(\beta)}=
\end{equation*}
\begin{equation*}
\frac{\alpha+\beta+1}{\alpha'+\beta +1}
C^{MN;B(m-1),D(m-1)}
(C^{\{k\};}{}_{MN}p^{\alpha'}s^{\beta}\xi^{\{n\}})_{B(m-1),D(m-1)}\Big|_{(\beta+1)th}.
\end{equation*}

\providecommand{\href}[2]{#2}\begingroup\raggedright\endgroup

\end{document}